\documentclass[12pt]{article}

\usepackage{amssymb}
\usepackage{amsmath}

\usepackage{graphicx}

\begin{document} %


\title{Polarized light-by-light scattering at the CLIC induced by axion-like particles}

\author{
S.C. \.{I}nan\thanks{Electronic address: sceminan@cumhuriyet.tr}
\\
{\small Department of Physics, Sivas Cumhuriyet University, 58140,
Sivas, Turkey}
\\
{\small and}
\\
A.V. Kisselev\thanks{Electronic address:
alexandre.kisselev@ihep.ru} \\
{\small Division of Theoretical Physics, A.A. Logunov Institute for
High Energy Physics,}
\\
{\small NRC ``Kurchatov Institute'', 142281, Protvino, Russia}}

\date{}

\maketitle

\begin{abstract}
The light-by-light (LBL) scattering with initial polarized Compton
backscattered photons at the CLIC, induced by axion-like particles
(ALPs), is investigated. The total cross sections are calculated,
assuming CP-even coupling of the pseudoscalar ALP to photons. The
95\% C.L. exclusion region for the ALP mass $m_a$ and its coupling
constant $f$ is presented. The results are compared with previously
obtained CLIC bounds for the unpolarized case. It is shown that the
bounds on $f$ for the polarized beams in the region $m_a = 1000
\mathrm{\ GeV} - 2000 \mathrm{\ GeV}$, with the collision energy
3000 GeV and integrated luminosity 4000 fb$^{-1}$, are on average
1.5 times stronger than the bounds for the unpolarized beams.
Herewith, our CLIC bounds are stronger than all current exclusion
regions for $m_a > 80$ GeV. In particular, they are more restrictive
than the limits which follow from the ALP-mediated LBL scattering at
the LHC.
\end{abstract}

\maketitle


\section{Introduction} %

The fine-tuning problem, known as the strong CP problem, is one of
the open issues of the Standard Model (SM). It can be solved by
introducing a spontaneously broken Peccei-Quinn symmetry
\cite{Peccei:1977_1,Peccei:1977_2} which involves a light
pseudoscalar particle, QCD axion \cite{Weiberg:1978,Wilzcek:1978}.
The QCD axion couples to the gluon field strength. Its phenomenology
is determined by its low mass and very weak interactions. In
particular, it could i) affect cosmology; ii) affect stellar
evolution; iii) mediate new long-range forces; iv) be produced in a
terrestrial laboratory. At present, the QCD axion is regarded as a
main component of the dark matter
\cite{Preskill:1983}-\cite{Dine:1983}. The solar axion
\cite{Bibber:1989} has been proposed to explain the excess in the
low-energy electron recoil observed by the XENON1T Collaboration
\cite{XENON}, since its energy spectrum matches the excess.

An axion-like particle (ALP) is a particle having interactions
similar to the axion. The origin of the ALP is expected to be
similar but without the relationship between its coupling constant
and mass. It means that the ALP mass can be treated independently of
its couplings to the SM fields. The ALPs emerge in string theory
scenarios \cite{Svrcek:2006}-\cite{Halverson:2019}, in theories with
spontaneously broken symmetries \cite{Masso:1997,Bellazzini:2017},
or in the GUT \cite{Rubakov:1997}. All these models predict an
ALP-photon coupling and, therefore, the electromagnetic decay of the
ALPs in two photons. Experimental searches are mainly directed to
ALPs, in order to relax the coupling parameter
\cite{Irastorza:2018}.

\begin{figure}[htb]
\begin{center}
\includegraphics[scale=0.8]{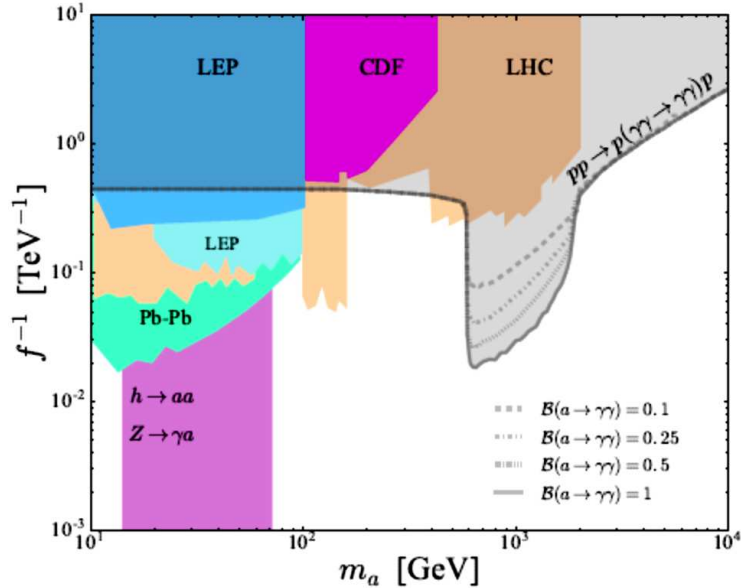}
\caption{The $95\%$ C.L. current exclusion regions for different
values of the ALP branching $\mathrm{Br}(a \rightarrow
\gamma\gamma)$. Here $f^{-1}$ is the ALP-photon coupling, and $m_a$
is the ALP mass. The figure is taken from
Ref.~\protect\cite{Beldenegro:2018}.} \label{fig:balfig7}
\end{center}
\end{figure}

The heavy ALPs can be detected at colliders in a light-by-light
(LBL) scattering \cite{Bauer:2017}-\cite{Beldenegro:2019}. It was
shown that LHC searches with the use of the proton tagging technique
constrain the ALP masses in the region 0.5 TeV -- 2 TeV
\cite{Beldenegro:2018}-\cite{Bauer:2019}. The current exclusion
regions for the axion and ALP searches are shown in
Fig.~\ref{fig:balfig7}.
%
The first evidence of the subprocess $\gamma\gamma \rightarrow
\gamma\gamma$ was observed by the ATLAS
\cite{ATLAS_ions_1,ATLAS_ions_2} and CMS \cite{CMS_ions}
Collaborations in high-energy ultra-peripheral PbPb collisions. The
phenomenological analysis of the exclusive and diffractive
$\gamma\gamma$ production in PbPb scattering at the LHC and FCC was
done in \cite{Coelho:2020_1,Coelho:2020_2}. The photon-induced
process $pp \rightarrow p\gamma\gamma p \rightarrow p'\gamma\gamma
p'$ at the LHC was studied in \cite{Atag:2009}-\cite{Inan:2019}.

We have recently investigated the virtual production of the ALPs in
the LBL scattering at the compact linear collider (CLIC)
\cite{Braun:2008,Boland:2016} with the initial unpolarized Compton
backscattered (CB) photons \cite{Inan:2020}. The 95\% C.L. exclusion
regions for the ALP mass $m_a$ and ALP-photon coupling $f$ have been
calculated. It turned out that our CLIC bounds on $m_a$ and $f$ are
stronger than the bounds for the LBL production of the ALP at the
LHC presented in Fig.~\ref{fig:balfig7}. Thus, the ALP search at the
CLIC has a great physics potential of searching for the ALPs,
especially in the mass region 1 TeV -- 2.4 TeV \cite{Inan:2020}.

The CLIC is planned to accelerate and collide electrons and
positrons at maximally 3 TeV center-of-mass energy. Three energy
states of the CLIC with $\sqrt{s} = 380$ GeV, $\sqrt{s} = 1500$ GeV
and $\sqrt{s} = 3000$ GeV are considered. The expected integrated
luminosities are $L=1000$ fb$^{-1}$, $L=2500$ fb$^{-1}$ and $L=5000$
fb$^{-1}$, respectively. At first two stages, it will be enable to
study the gauge sector, Higgs and top physics with a high precision.
At the third stage, the most precise investigation of the SM, as
well as new physics will be possible \cite{Dannheim:2012,CLIC_BSM}.

At the CLIC, it is possible to study not only $e^+e^-$ scattering
but also $\gamma\gamma$ collisions with real photons. These photon
beams are given by the Compton backscattering of laser photons off
linear electron beams. The physics potential of a linear collider is
greatly enhanced with polarized beams \cite{polarized_beams}. The SM
backgrounds may be reduced by a factor of five if the electron beam
has a polarization of 80\%. Searches for new physics can also be
enhanced when using polarization beams. The CLIC accelerator
conceptual design includes a source to produce a polarized electron
beam and all elements necessary to transport the beam to the IP
without loss of polarization. An electron beam polarization of 80\%
is expected for the baseline CLIC experimental programme.

In our recent paper \cite{Inan:2020}, the axion induced LBL
scattering of the \emph{unpolarized} CB photons was investigated. In
the present paper, we propose to study the same process with ingoing
\emph{polarized} CB photon beams. A summation over outgoing photons
is assumed. The main goal is to demonstrate that the CLIC bounds on
the ALP parameters can be improved if one considers the polarized
LBL scattering.

\section{Polarized real photon beams} %

As was already mentioned above, $\gamma\gamma$-interactions with
real photons can be examined at the CLIC. Real photons beams are
obtained by the Compton backscattering of laser photons off linear
electron beams. Most of these real scattered photons have high
energy, and the $\gamma\gamma$ luminosity turns out to be of the same
order as the one for $e^+e^-$ collision \cite{Ginzburg:1981}. That
is why one gets a large cross section for the LBL scattering of the
real photons.

The spectrum of backscattered photons is given by helicities of
initial laser photon and electron beam as follows
\begin{align}
f_{\gamma/e}(y) = {{1}\over{g(\zeta)}} & \Big[ 1-y + {{1}\over{1-y}}
-{{4y}\over{\zeta(1-y)}}+{{4y^{2}}\over {\zeta^{2}(1-y)^{2}}}
\nonumber \\
&+ \lambda_{0}\lambda_{e} r\zeta (1-2r)(2-y) \Big]
\label{photon_spectrum} ,
\end{align}
where
\begin{align}
g(\zeta) &= g_{1}(\zeta)+
\lambda_{0}\lambda_{e} \,g_{2}(\zeta) \;, \label{g_zeta} \\
g_{1}(\zeta) &= \left( 1-{{4}\over{\zeta}} - {{8}\over{\zeta^{2}}}
\right) \ln{(\zeta+1)}
+ {{1}\over{2}}+{{8}\over{\zeta}} - {{1}\over{2(\zeta+1)^{2}}} \;, \label{g_zeta1} \\
g_{2}(\zeta) &= \left( 1 + {{2}\over{\zeta}} \right) \ln{(\zeta+1)}
- {{5}\over{2}}+{{1}\over{\zeta+1}} - {{1}\over{2(\zeta+1)^{2}}}
\label{g_zeta2} \;,
\end{align}
and
\begin{equation}\label{zeta_y_r}
\zeta = \frac{4E_{e}E_{0}}{M_{e}^{2}} \;, \quad  y =
\frac{E_{\gamma}}{E_{e}} \;, \quad r = \frac{y}{\zeta(1-y)}  \;.
\end{equation}
Here $E_\gamma$ is the scattered photon energy, $E_0$ and
$\lambda_0$ are the energy, and the helicity of the initial laser photon
beam, $E_e$, and $\lambda_e$ are the energy and helicity of the
initial electron beam before CB. Note that the variable $y$ reaches
its maximum value 0.83 when $\zeta = 4.8$. The helicity of the CB
photons,
\begin{eqnarray}
\xi(E_{\gamma},\lambda_{0}) = {{\lambda_{0}(1-2r)
(1-y+1/(1-y))+\lambda_{e} r\zeta[1+(1-y)(1-2r)^{2}]}
\over{1-y+1/(1-y)-4r(1-r)-\lambda_{e}\lambda_{0}r\zeta (2r-1)(2-y)}}
\;,
\end{eqnarray}
has the highest value when $y \simeq 0.83$. In what follows, we will
consider two cases:
\begin{align}\label{}
(\lambda_0^{(1)}, \lambda_e^{(1)}; \lambda_0^{(2)}, \lambda_e^{(2)})
&= (1, -0.8; 1,-0.8) \;, \nonumber \\
(\lambda_0^{(1)}, \lambda_e^{(1)}; \lambda_0^{(2)}, \lambda_e^{(2)})
&= (1, +0.8; 1,+0.8) \;,
\end{align}
where the superscripts 1 and 2 enumerate the beams. As for the
integrated luminosities for the baseline CLIC energy stages, we will
take them from Ref.~\cite{CLIC_lum}, see Tab.~1.

\begin{center}
\begin{tabular}{||c|c||c||c|c||}
  \hline
\multicolumn{2}{||c||} {} & {Unpolarized}  & $\lambda_e = -0.8$ & $\lambda_e = +0.8$ \\
\hline
 Stage & $\sqrt{s}$, GeV & $\mathcal{L}$, fb$^{-1}$ & $\mathcal{L}$, fb$^{-1}$ &  $\mathcal{L}$, fb$^{-1}$ \\
 \hline
  1 & 380 & 1000 & 500 & 500 \\
  2 & 1500 & 2500 & 2000 & 500 \\
  3 & 3000 & 5000 & 4000 & 1000 \\
  \hline
\end{tabular}
\end{center}

\begin{center}
Table~1. The CLIC energy stages and integrated luminosities for the
unpolarized and polarized electron beams.
\end{center}
As one can see from Tab.~1, for the polarized electron beams the
luminosities are noticeably smaller than those for the unpolarized
beams, especially for the first two energy stages and $\lambda_e =
0.8$.

Numerical estimates have shown that for  $\sqrt{s} = 380$ GeV the
total cross sections almost coincide with the SM cross sections
\cite{Inan:2020}. That is why, we will do our calculations for the
collision energies $\sqrt{s} = 1500$ GeV (2nd stage of the CLIC) and
$\sqrt{s} = 3000$ GeV (3rd stage of the CLIC).

\section{Light-by-light production of ALP} %

We will consider a Lagrangian with the CP-even coupling of the
pseudoscalar ALP (in what follows, denoted as $a$) to photons, and
the ALP coupling to fermions,
\begin{equation}\label{axion_photon_lagrangian}
\mathcal{L}_a = \frac{1}{2}\,\partial_\mu a \,\partial^{\,\mu} \!a -
\frac{1}{2} m_a^2 a^2 + \frac{a}{f} \,F_{\mu\nu} \tilde{F}^{\mu\nu}
+ \frac{\partial_\mu a}{2f} \sum_{\psi} c_\psi \bar{\psi} \gamma^\mu
\psi \;,
\end{equation}
where $F_{\mu\nu}$ is the electromagnetic tensor,
$\tilde{F}_{\mu\nu} = (1/2) \varepsilon_{\mu\nu\rho\sigma}
F^{\rho\sigma}$ its dual, $c_\psi$ is a dimensionless constant. Note
that, contrary to the QCD axion, the ALP does not couple to the
gluon anomaly. The ALP-photon coupling $f$ defines the ALP decay
width into two photons
\begin{equation}\label{photon_axion_width}
\Gamma(a\rightarrow\gamma\gamma) = \frac{m_a^3}{4\pi f^2} \;,
\end{equation}
and the decay rate of the ALP to fermions,
\begin{equation}\label{lepton_axion_width}
\Gamma(a \rightarrow \bar{\psi} \psi) = \frac{m_a  m_\psi^2}{8\pi}
\left( \frac{c_\psi}{f} \right)^{\!\!2} \sqrt{1 -
\frac{4m_\psi^2}{m_a^2}} \;,
\end{equation}
where $m_\psi$ is the fermion mass. As one can see from
eqs.~\eqref{photon_axion_width}, \eqref{lepton_axion_width}, for
$m_a \gg m_\psi$, and $c_\psi = \mathrm{O}(1)$ the full width of the
ALP will be mainly defined by its decay into two photons. In
general, the ALP branching $\mathrm{Br}(a\rightarrow\gamma\gamma)$
can be less than 1.

The differential cross section of the diphoton production with the
initial polarized CB photons is defined by \cite{Cakir:2008}
\begin{align}\label{diff_cs}
\frac{d\sigma}{d\cos \theta} &= \frac{1}{128\pi s} \int\limits_{x_{1
\min}}^{0.83} \!\!\frac{dx_1}{x_1} \,f_{\gamma/e}(x_1)
\int\limits_{x_{2 \min}}^{0.83}
\!\!\frac{dx_2}{x_2} \,f_{\gamma/e}(x_2) \nonumber \\
&\times \Big\{ \left[ 1 + \xi \left(
E_{\gamma}^{(1)},\lambda_{0}^{(1)} \right) \xi \left(
E_{\gamma}^{(2)},\lambda_{0}^{(2)} \right) \right]
|M_{++}|^2 \nonumber \\
&\quad + \left[ 1 -\xi \left( E_{\gamma}^{(1)},\lambda_{0}^{(1)}
\right) \xi \left( E_{\gamma}^{(2)},\lambda_{0}^{(2)} \right)
\right] |M_{+-}|^2 \Big\} ,
\end{align}
where $x_i = E_{\gamma}^{(i)}/E_e$ ($i=1,2$) are the energy
fractions of the CB photon beams, $x_{1 \min} = p_\bot^2/E_e^2$,
$x_{2 \min} = p_\bot^2/(x_{1} E_e^2)$, $p_{\bot}$ is the transverse
momentum of the final photons. Here $\sqrt{s}$ is the center of mass
energy of the $e^+e^-$ collider, while $\sqrt{s x_1 x_2}$ is the
center of mass energy of the backscattered photons. The amplitudes
$|M_{++}|$ and $|M_{+-}|$ are obtained by summations over the
helicities of the outgoing photons in the helicity amplitudes,
\begin{align}\label{helicity_sum}
|M_{++}|^2 &= |M_{++++}|^2 + |M_{++--}|^2 \;, \nonumber \\
|M_{+-}|^2 &= |M_{+-+-}|^2 + |M_{+--+}|^2 \;.
\end{align}
We have used P-, T-, and Bose symmetries. In its turn, each of the
amplitudes is a sum of the ALP and SM terms,
\begin{equation}\label{M_tot}
M = M_a + M_{\mathrm{SM}} \;.
\end{equation}
As the main SM background, we will take into account both $W$-loop
and fermion-loop contributions
\begin{equation}\label{M_ew}
M_{\mathrm{SM}} = M_f + M_W \;.
\end{equation}
The explicit analytical expressions for SM helicity amplitudes in
the right-hand side of eq.~\eqref{helicity_sum}, both for the
fermion and W-boson terms, are too long. That is why we do not
present them here. They can be found in \cite{Inan:2020} (see also
\cite{Beldenegro:2018}). In order to reduce the SM background, we
will impose the cut on a rapidity of the final state photons,
$|\eta_{\gamma\gamma}| < 2.5$. Finally, a possible background with
fake photons from decays of $\pi^0$, $\eta$, and $\eta'$ is
negligible in the signal region.

\begin{figure}[htb]
\begin{center}
\includegraphics[scale=0.6]{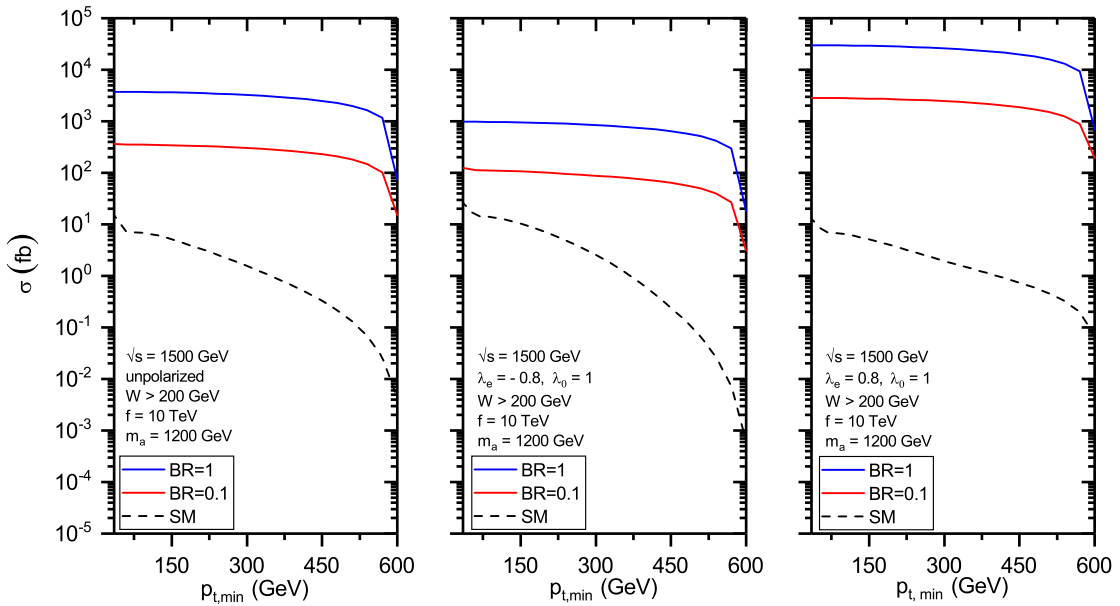}
\caption{The total cross sections for the process $\gamma\gamma
\rightarrow \gamma\gamma$ at the CLIC as functions of the transverse
momenta cutoff $p_{\mathrm{t,min}}$ of the final photons for the
invariant energy $\sqrt{s} = 1500$ GeV. Left panel: unpolarized
case. Middle panel: the helicity of the electron beam is $\lambda_e
= -0.8$. Right panel: the helicity of the electron beam is
$\lambda_e = 0.8$.} \label{fig:CBPTCUTE750C}
\end{center}
\end{figure}

\begin{figure}[htb]
\begin{center}
\includegraphics[scale=0.6]{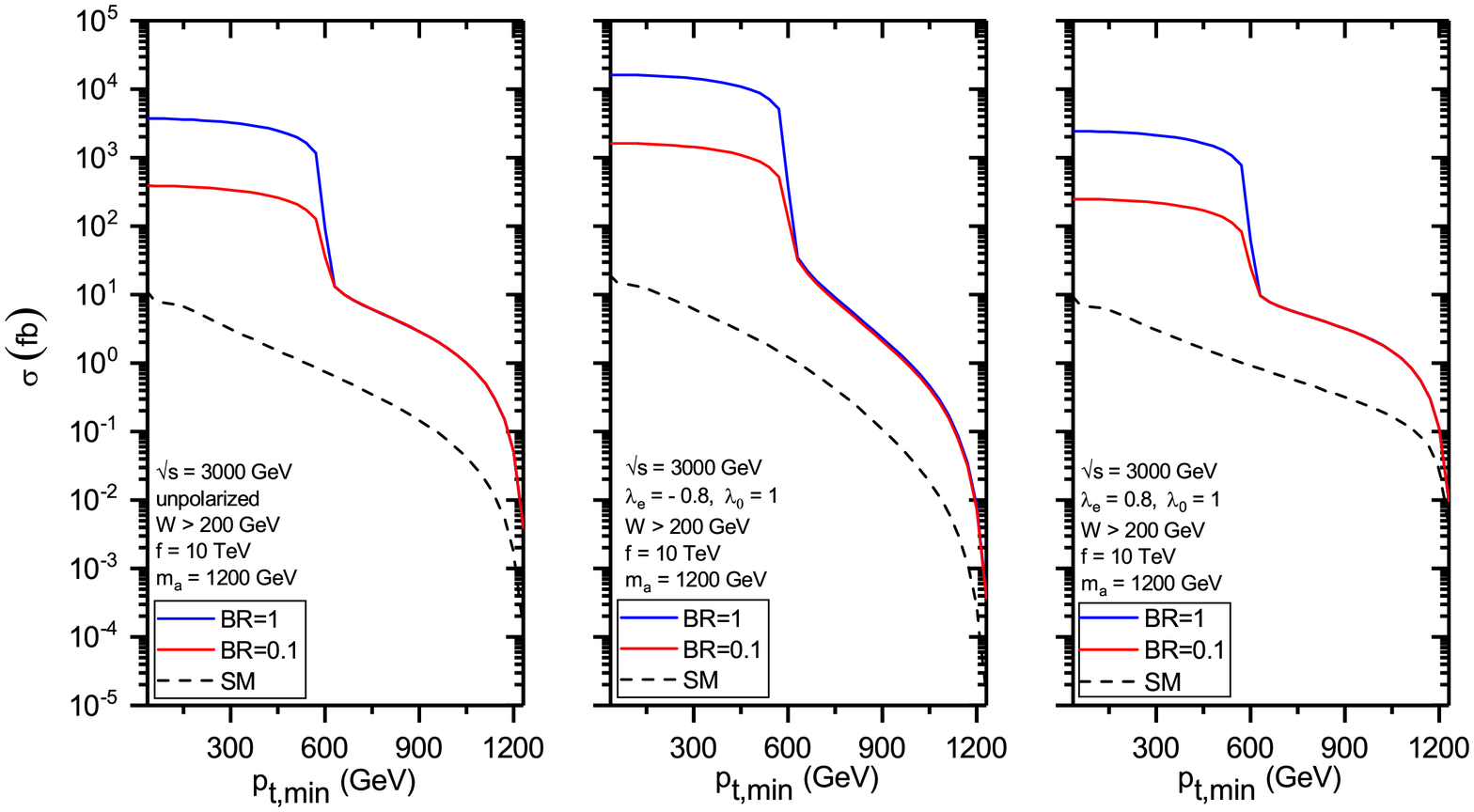}
\caption{The total cross sections for the process $\gamma\gamma
\rightarrow \gamma\gamma$ at the CLIC as functions of the transverse
momenta cutoff $p_{\mathrm{t,min}}$ of the final photons for the
invariant energy $\sqrt{s} = 3000$ GeV. Left panel: unpolarized
case. Middle panel: the helicity of the electron beam is $\lambda_e
= -0.8$. Right panel: the helicity of the electron beam is
$\lambda_e = 0.8$.} \label{fig:CBPTCUTE1500C}
\end{center}
\end{figure}

In Figs.~\ref{fig:CBPTCUTE750C} and \ref{fig:CBPTCUTE1500C} the
total cross sections for the process $\gamma\gamma \rightarrow
\gamma\gamma$ with the unpolarized and polarized CB initial photons
are shown as functions of the minimal transverse momenta of the
final photons $p_{t,\min}$. In Fig.~\ref{fig:CBPTCUTE750C} the
invariant energy is taken to be $\sqrt{s} = 1500$ GeV, the ALP mass
$m_a$ and its coupling $f$ are chosen to be equal to 1200 GeV and 10
TeV, respectively. In order to reduce the SM background, we have
imposed the cut on the invariant energy of the final photons $W =
m_{\gamma\gamma} > 200$ GeV. The cross sections are presented for
two values of the ALP branching $\mathrm{Br} = \mathrm{Br}(a
\rightarrow \gamma\gamma)$. The curves in the left, middle and right
panels correspond to the helicity of the initial electron beam
before CB $\lambda_{e} = 0$ (unpolarized case), $\lambda_{e} =
-0.8$, and $\lambda_{e} = 0.8$, respectively. The SM predictions are
also presented.  The total cross sections for $\sqrt{s} = 3000$ GeV
are shown in Fig.~\ref{fig:CBPTCUTE1500C}. As one can see, the
deviation from the SM gets higher as $p_{t,\min}$ increases,
especially for $\lambda_{e} = 0.8 \ (-0.8)$, if $\sqrt{s} = 1500 \
(3000)$ GeV. Let us note that for $\sqrt{s} = 1500$ GeV,
$\lambda_{e} = -0.8$, the total cross section for the polarized
beams is even less that the unpolarized total cross section. The
same is true for $\sqrt{s} = 3000$ GeV and $\lambda_{e} = 0.8$.

\begin{figure}[htb]
\begin{center}
\includegraphics[scale=0.65]{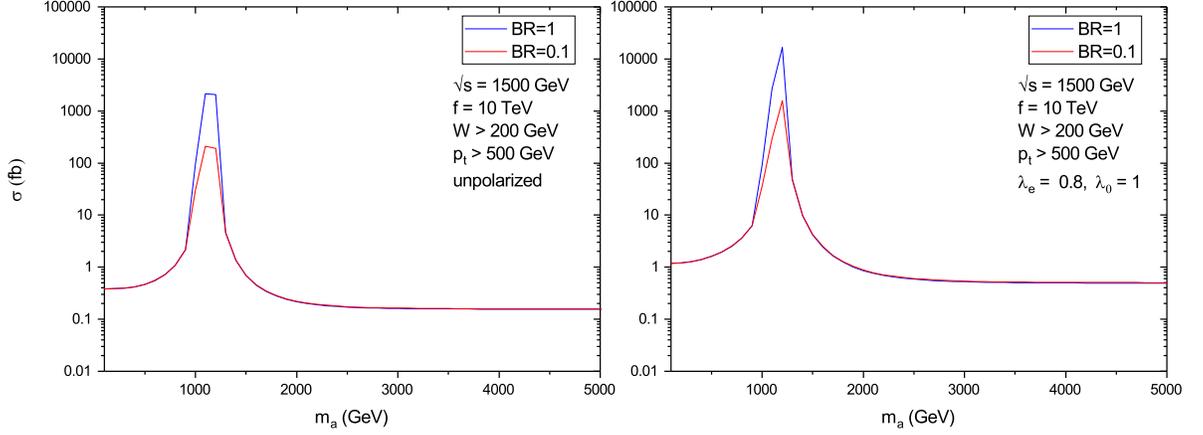}
\caption{The total cross sections for the process $\gamma\gamma
\rightarrow \gamma\gamma$ at the CLIC for the CB initial photons as
functions of the ALP mass $m_a$ for $\sqrt{s} = 1500$ GeV and $f =
10$ TeV. Left panel: unpolarized case. Right panel: polarized case,
the helicity of the electron beam is equal to $\lambda_e = 0.8$.}
\label{fig:CBMAF10_P}
\end{center}
\end{figure}

\begin{figure}[htb]
\begin{center}
\includegraphics[scale=0.65]{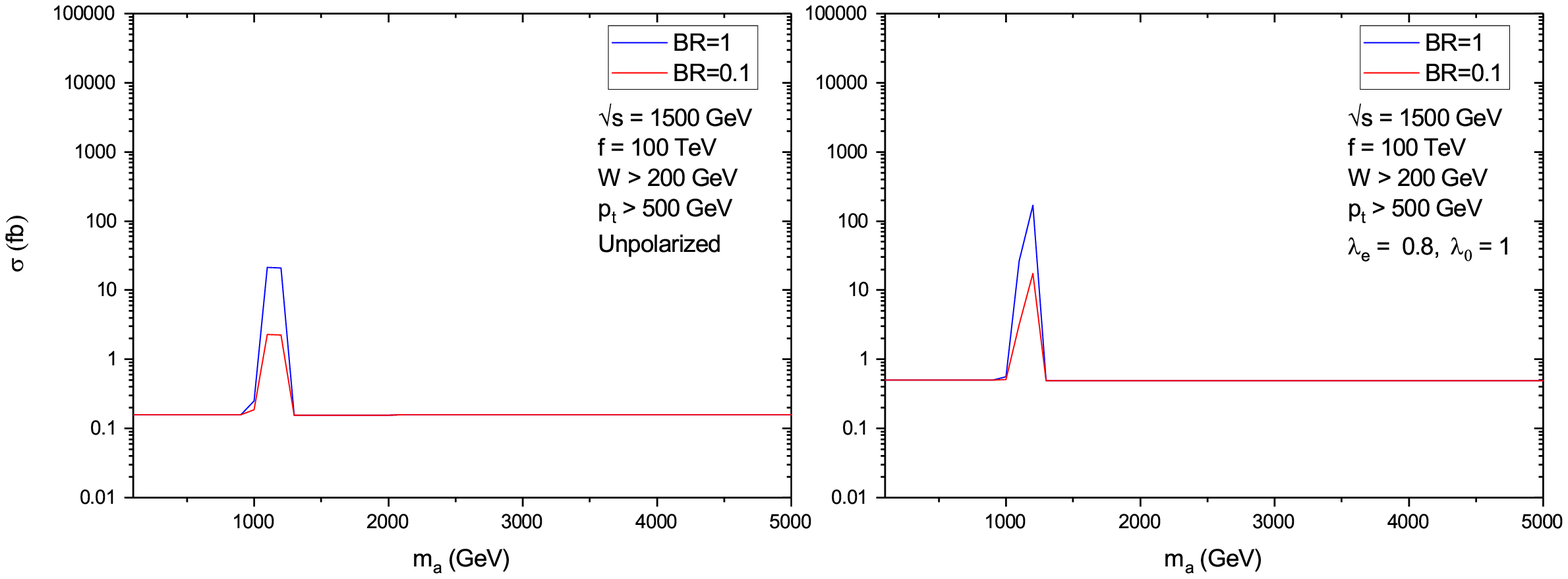}
\caption{The same as in Fig.~\ref{fig:CBMAF10_P}, but for $f = 100$
TeV.} \label{fig:CBMAF100_P}
\end{center}
\end{figure}

\begin{figure}[htb]
\begin{center}
\includegraphics[scale=0.65]{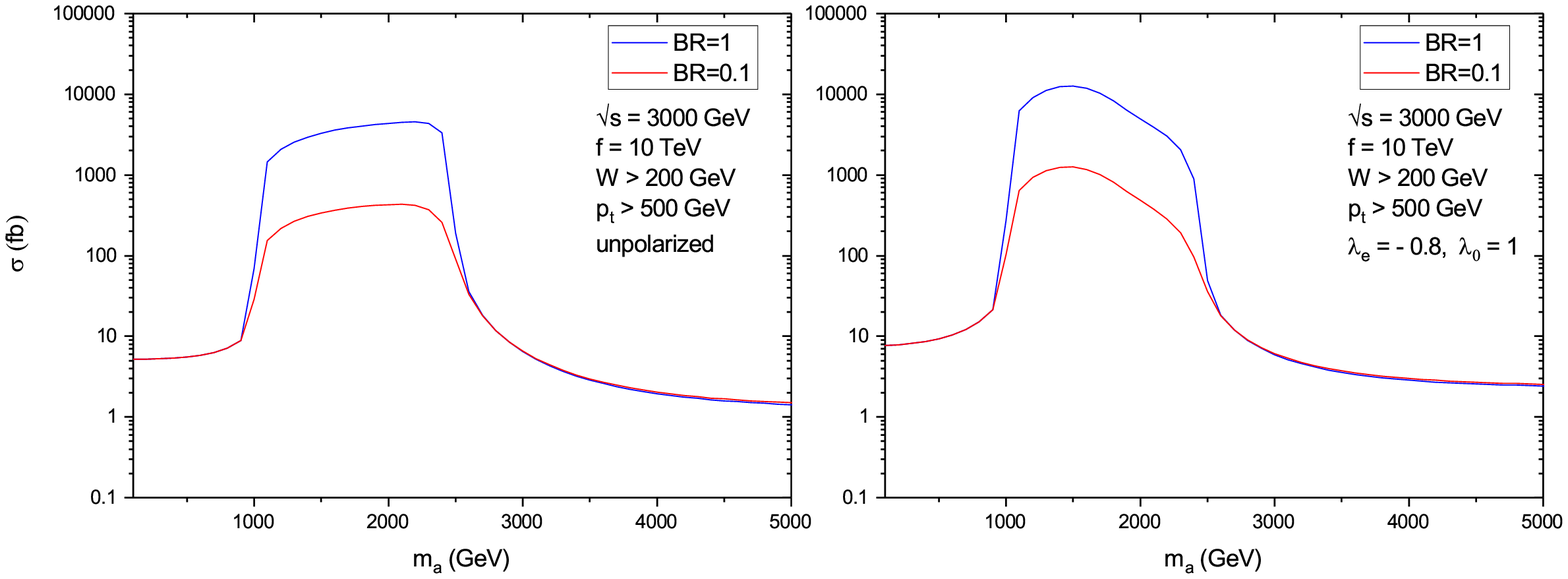}
\caption{The total cross sections for the process $\gamma\gamma
\rightarrow \gamma\gamma$ at the CLIC for the CB initial photons as
functions of the ALP mass $m_a$ for $\sqrt{s} = 3000$ GeV and $f =
10$ TeV. Left panel: unpolarized case. Right panel: polarized case,
the helicity of the electron beam is equal to $\lambda_e = - 0.8$.}
\label{fig:CBMAF10}
\end{center}
\end{figure}

\begin{figure}[htb]
\begin{center}
\includegraphics[scale=0.65]{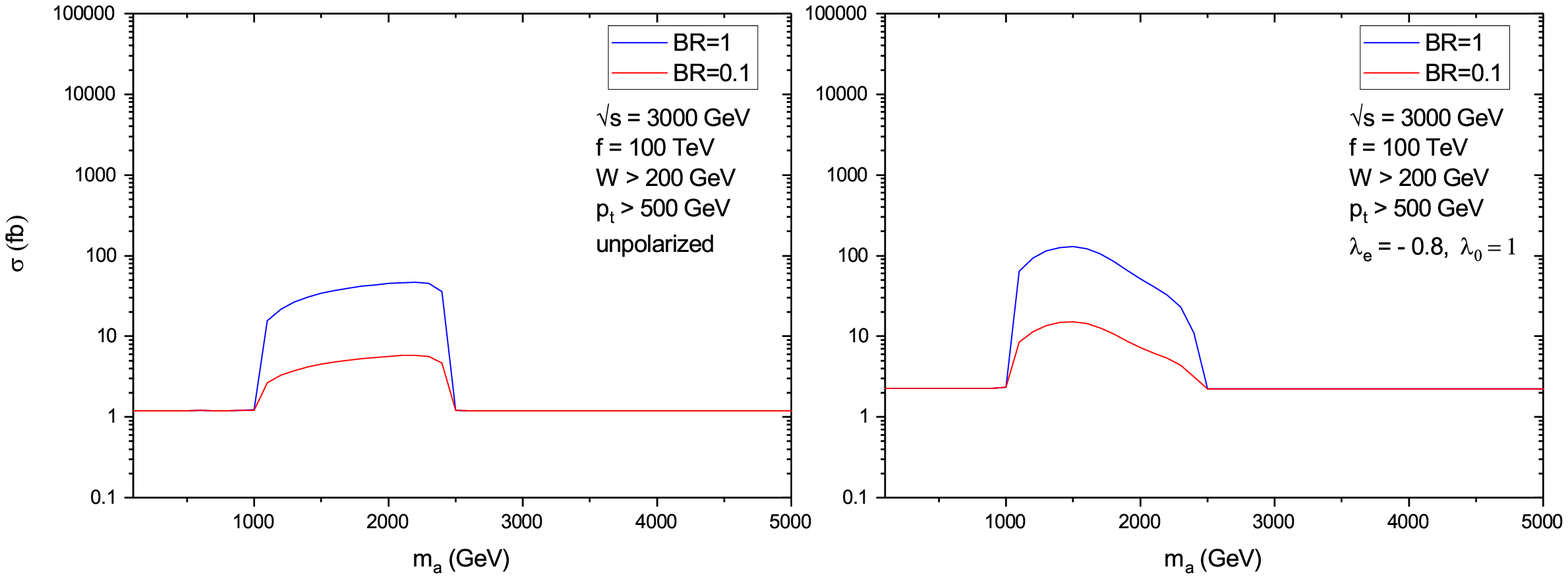}
\caption{The same as in Fig.~\ref{fig:CBMAF10}, but for $f = 100$
TeV.} \label{fig:CBMAF100}
\end{center}
\end{figure}

Figs.~\ref{fig:CBMAF10_P} and \ref{fig:CBMAF100_P} demonstrate the
dependence of the total cross sections on the ALP mass both for
unpolarized and polarized electron beams for $\sqrt{s} = 1500$ GeV,
two values of the coupling constant $f$, and two values of the ALP
branching $\mathrm{Br(a \rightarrow \gamma\gamma)}$. The total cross
sections for $\sqrt{s} = 3000$ GeV are shown in
Figs.~\ref{fig:CBMAF10} and \ref{fig:CBMAF100}. All the curves have
sharp peaks near point $m_a = 1200$ GeV. As mentioned after
eq.~\eqref{zeta_y_r}, the maximum value of the ratio $E_\gamma/E_e$
is equal to 0.83. That is why, a bump around $0.83 \times 1500$ GeV
= 1245 GeV should be expected, in agreement with
Figs.~\ref{fig:CBMAF10_P}, \ref{fig:CBMAF100_P}. As one can see, for
$\sqrt{s} = 1500$ GeV the polarized cross sections exceed the
unpolarized cross sections by an order of magnitude. Unfortunately,
due to the relatively small integrated luminosity for the second
CLIC stage (see Tab.~1), expected bounds on $m_a$ and $f$ appears to
be even less stronger than corresponding bounds for the unpolarized
case. Thus, we have to deal with the third energy stage of the CLIC.
For $\sqrt{s} = 3000$ GeV the ratio of the polarized cross section
to unpolarized one is approximately equal to 2.5.

One can see that the cross sections in Figs.~\ref{fig:CBMAF10},
\ref{fig:CBMAF100}  are very sensitive to the parameter $m_a$ in the
interval $m_a = 1000-2500$ GeV, in which it is approximately two
orders of magnitude greater than for $m_a$ outside of this mass
region. An approximate formula for the cross section with the CB
initial photons can be obtained, which explains a non-trivial
dependence of the cross section on the ALP mass $m_a$, its coupling
constant $f$ and $\mathrm{Br}(a\rightarrow\gamma\gamma)$ in the mass
region $1000-2500$ GeV. The point is that a dominant contribution to
the cross section comes from $s$-channel terms in the matrix element
$M$. To illustrate this point, let us put
\begin{equation}\label{M}
M = \frac{4}{f^2} \frac{s^2}{s - m_a^2 + i m_a\Gamma_a} \;.
\end{equation}
The calculations show that the most important energy region is a
resonance region $s \sim m_a^2$ in which
\begin{equation}\label{M2_reson}
|M|^2 \big|_{s \sim m_a^2} \sim \frac{m_a^6}{f^4 \Gamma_a^2} \;.
\end{equation}
Since our matrix element \eqref{M} depends only on $s$, the cross
section of the subprocess $\gamma\gamma\rightarrow\gamma\gamma$  is
given by the integral
\begin{equation}\label{proc_cs}
\sigma = \frac{1}{4E_e^2}\int \!\!ds \,\frac{1}{16\pi s} \,|M|^2 \;.
\end{equation}
Let us estimate the contribution to $\sigma$ from the resonance
region
\begin{equation}\label{region}
m_a^2 - C m_a \Gamma_a \leqslant s \leqslant m_a^2 + C m_a \Gamma_a
\;,
\end{equation}
where $C$ is a constant of order $\mathrm{O}(1)$. Then we get
\begin{equation}\label{cross_section_2}
\sigma = \frac{1}{f^2} \,\frac{m_a^2}{E_e^2}
\,\mathrm{Br}(a\rightarrow\gamma\gamma)\int_{-C}^{C} dx
\,\frac{1}{x^2 + 1} \;.
\end{equation}
As a result, we find for $C=1$ that
\begin{equation}\label{cs_final}
\sigma \simeq 0.6 \left( \frac{\mathrm{TeV}}{f} \right)^{\!2} \left(
\frac{m_a}{E_e}\right)^{\!2} \mathrm{Br}(a\rightarrow\gamma\gamma)
\mathrm{\ fb} \;.
\end{equation}
This formula gives a correct dependence of the cross section on the
parameters $f$, $m_a$ and $\mathrm{Br}(a\rightarrow\gamma\gamma)$ in
the mass region $1100-2500$ GeV, see Figs.~\ref{fig:CBMAF10} and
\ref{fig:CBMAF100}. Note that $\sigma$ is proportional to $1/f^2$
\eqref{cs_final}, while simple dimensional arguments would give us
$1/f^4$ dependence. Let us underline that the above considerations
are not applicable outside the mass region $1100-2500$ GeV.

\begin{figure}[htb]
\begin{center}
\includegraphics[scale=0.65]{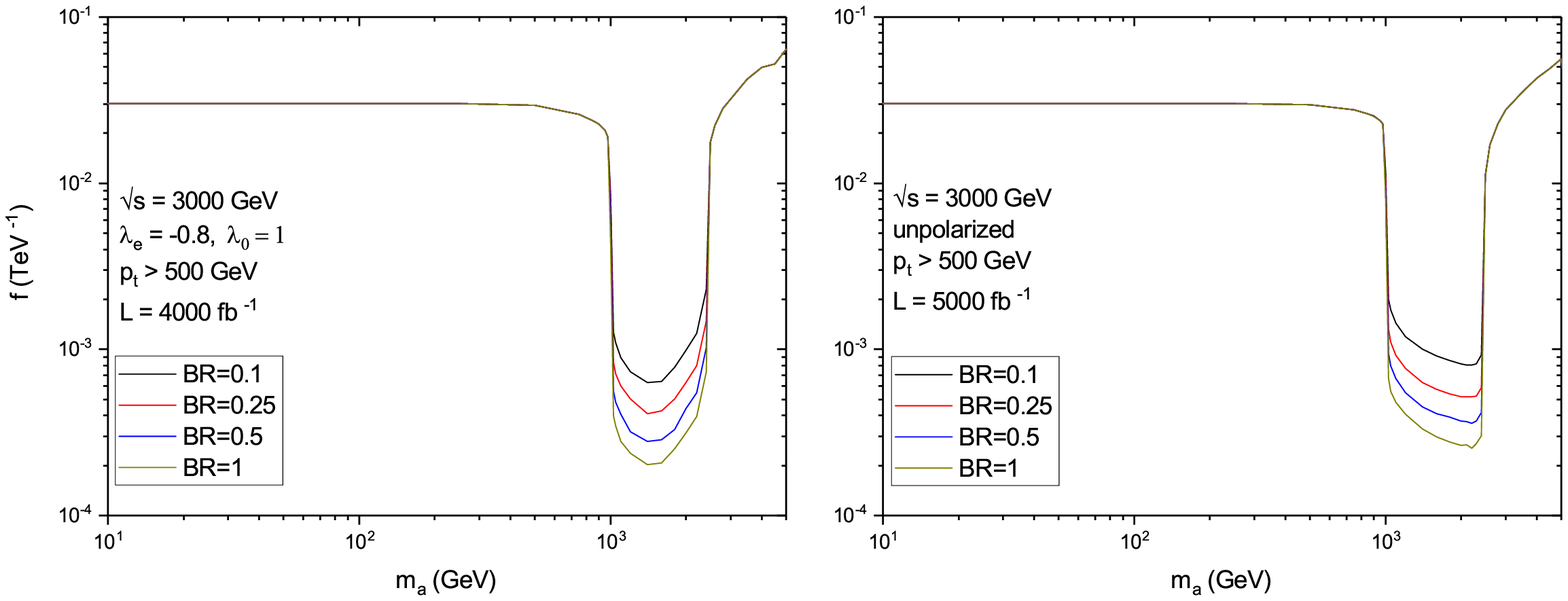}
\caption{The 95\% C.L. CLIC exclusion region for the process
$\gamma\gamma \rightarrow \gamma\gamma$ with the CB ingoing photons
and invariant energy $\sqrt{s} = 3000$ GeV. Left panel: polarized
electron beams with the helicity $\lambda_e = -0.8$; integrated
luminosity $L = 4000$ fb$^{-1}$. Right panel:  unpolarized electron
beams; integrated luminosity $L = 5000$ fb$^{-1}$
\protect\cite{Inan:2020}. } \label{fig:CBSS}
\end{center}
\end{figure}

As already mentioned above, the cross sections are very sensitive to
the parameter $m_a$ in the interval $m_a = 1000-2500$ GeV, in which
it is approximately two orders of magnitude greater than for $m_a$
outside of this mass region, see Figs.~\ref{fig:CBMAF10},
\ref{fig:CBMAF100}. It is not surprising that this is the region
where the value of the ALP coupling constant $f$ is mostly
restricted by the polarized LBL process. The exclusion region is
presented in the left panel of Fig.~\ref{fig:CBSS} in comparison
with the unpolarized case is shown in the right panel of this
figure. We have used the following formula for calculating the
statistical significance ($SS$) \cite{SS}
\begin{equation}\label{SS_def}
SS = \sqrt{2[(S+B) \,\ln(1 + S/B) - S]} \;,
\end{equation}
where $S$ and $B$ are the numbers of the signal and background
events, respectively. It was assumed that the uncertainty of the
background is negligible. In order to suppress the SM background, we
have applied the cut $p_t > 500$ GeV on the momenta of the final
photons.

As it follows from Fig.~\ref{fig:CBSS}, the best bounds for the LBL
scattering at the CLIC are realized for $\mathrm{Br(a \rightarrow
\gamma\gamma)} = 1$. Herewith, we have:
\begin{itemize}
  \item
For the mass region $10 \mathrm{\ GeV} < m_a < 500$ GeV, the
polarized and unpolarized upper bounds on $f$ are almost the same,
$f^{-1} = 3.0 \times 10^{-2}$ TeV$^{-1}$.
  \item
In the interval $500 \mathrm{\ GeV} < m_a < 1000$ GeV, the polarized
bounds are about 1.1 times better than the unpolarized ones. For
example, for $m_a= 850$ GeV, we find $f^{-1} = 2.65 \times 10^{-2}$
TeV$^{-1}$ for the unpolarized case, and $f^{-1} = 2.40 \times
10^{-2}$ TeV$^{-1}$ for the polarized case.
  \item
The region $1000 \mathrm{\ GeV} < m_a < 2000$ GeV is the best region
in which the polarized bounds are on average 1.5 times stronger. For
example, for  $m_a = 1400$ GeV,  $f^{-1} = 3.35 \times 10^{-4}$
TeV$^{-1}$ for the unpolarized beams, and $f^{-1} = 2.05 \times
10^{-4}$ TeV$^{-1}$ for the polarized beams.
  \item
In the mass interval $2000 \mathrm{\ GeV} < m_a < 2500$ GeV, the
unpolarized bounds are 2 times better on average. For instance, for
the $m_a = 2400$ GeV, we get $f^{-1} = 3.05 \times 10^{-4}$
TeV$^{-1}$, and $f^{-1} = 7.35 \times 10^{-4}$ TeV$^{-1}$ for the
unpolarized and polarized beams, respectively.
  \item
Finally, for $2500 \mathrm{\ GeV} < m_a < 5000$ GeV the unpolarized
bounds are 1.2 times better on average. In particular, for $m_a =
3500$ GeV, we find $f^{-1} = 3.35 \times 10^{-2}$ TeV$^{-1}$ for the
unpolarized beams, while $f^{-1} = 4.20 \times 10^{-2}$ TeV$^{-1}$
for the polarized beams.
\end{itemize}

\section{Conclusions} %

In the present paper, the light-by-light scattering with the ingoing
\emph{polarized} Compton backscattered photons at the CLIC, induced
by the axion-like particles has been studied. The total cross
sections are calculated for the $e^+e^-$ collider energies 1500 GeV
and 3000 GeV. The cross sections are presented as functions of the
ALP mass $m_a$, its coupling constant $f$, and ALP branching into
two photons $\mathrm{Br}(a \rightarrow \gamma\gamma)$. By combining
the results obtained with the results on the \emph{unpolarized}
light-by-light scattering derived recently in Ref.~\cite{Inan:2020},
we have to make the following conclusions:

\begin{enumerate}
  \item
First energy stage of the CLIC ($\sqrt{s} = 380$ GeV): \\
The SM contribution completely dominates the axion induced
contribution for $f=10$ TeV in the mass interval $m_a = 10 - 5000$
GeV. Any search of the ALPs is thus meaningless in this mass region.
  \item
Second energy stage of the CLIC ($\sqrt{s} = 1500$ GeV): \\
The axion contribution dominates the SM one both for the unpolarized
and polarized ingoing CB photons. For the electron beam helicity
$\lambda_e = -0.8$, the cross section is even smaller than the
unpolarized cross section, compare the middle and left panels of
Fig.~\ref{fig:CBPTCUTE750C}. But for $\lambda_e = 0.8$ the polarized
cross section exceeds the unpolarized one by order of magnitude, see
the right panel of Fig.~\ref{fig:CBPTCUTE750C}. Nevertheless, due to
the relatively small value of the expected integrated luminosity in
such a case (500 fb$^{-1}$, as compared with 2500 fb$^{-1}$ for the
unpolarized electron beams), the bounds on $m_a$ and $f$ are less
stronger than analogous bounds for the unpolarized LBL collision.
Thus, one has no advantages to use the polarized electron beams in
searching for heavy ALPs at this energy.
  \item
Third energy stage of the CLIC ($\sqrt{s} = 3000$ GeV): \\
For the electron beam helicity $\lambda_e = 0.8$ (right panel of
Fig.~\ref{fig:CBPTCUTE1500C}), the cross section is smaller than the
unpolarized cross section (left panel of
Fig.~\ref{fig:CBPTCUTE1500C}). However, for $\lambda_e = -0.8$ the
polarized cross section exceeds the unpolarized cross section by a
factor of 2.5, as one can see by comparing the middle and left
panels of this figure. Fig.~\ref{fig:CBSS} demonstrates us that the
bounds on $m_a$ and $f$ are better than recently obtained limits for
the unpolarized LBL collision in the mass region $m_a = 500
\mathrm{\ GeV} - 2000$ GeV. Especially, it takes place in the
interval $m_a = 1000 \mathrm{\ GeV} - 2000$ GeV in which the bounds
on $f$ for the polarized beams are on average 1.5 times stronger
than the bounds obtained for unpolarized beams.
\end{enumerate}

\begin{figure}[htb]
\begin{center}
\includegraphics[scale=0.2]{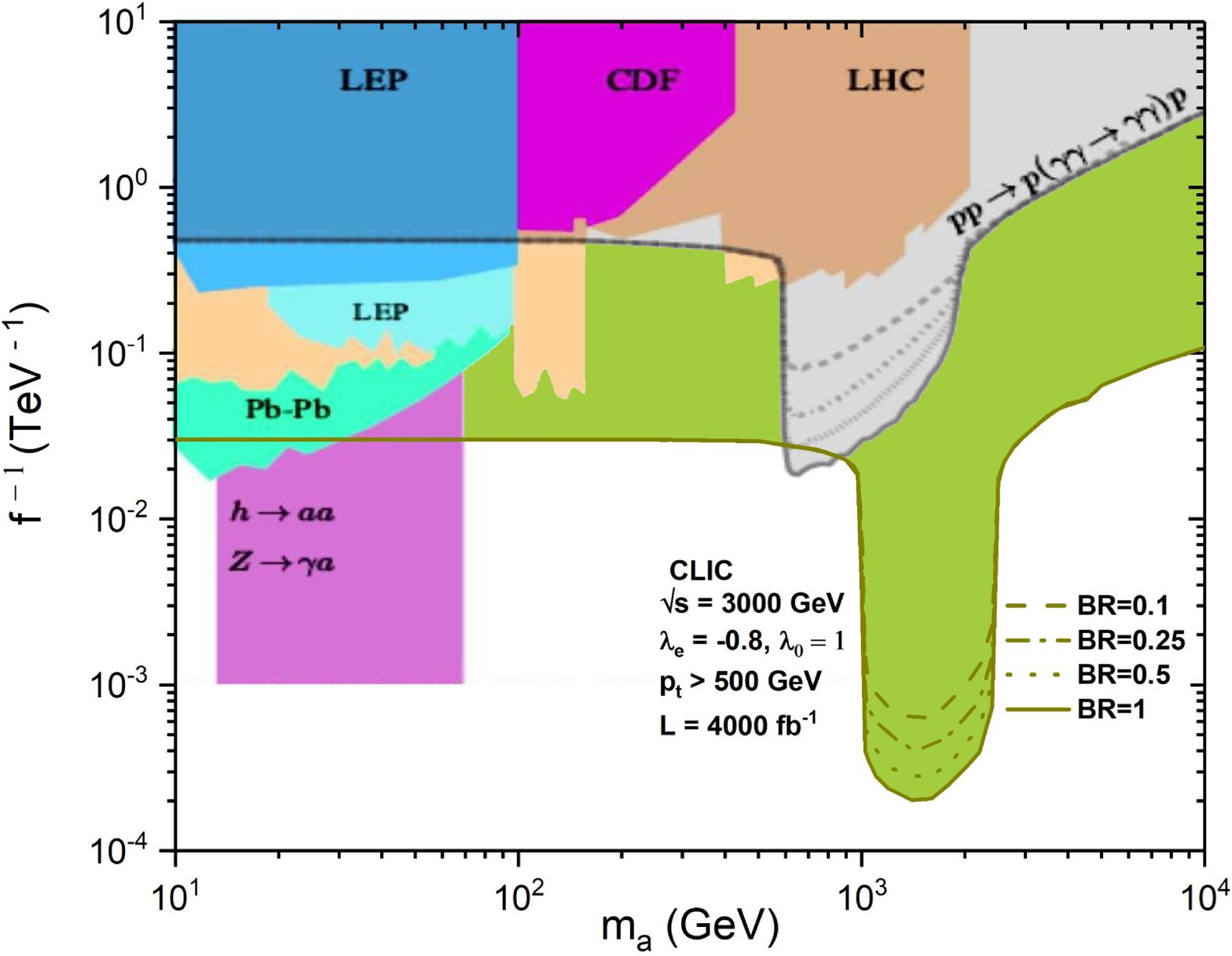}
\caption{Our prediction for the  $95\%$ C.L. exclusion region for
the energy $\sqrt{s} = 3000$ GeV, electron beam polarization
$\lambda_e = -0.8$, and different values of the ALP branching
$\mathrm{Br}(a \rightarrow \gamma\gamma)$ (green area), in
comparison with other current exclusion regions.}
\label{fig:CBSSPOL_all_f}
\end{center}
\end{figure}

Our main results are presented in Fig.~\ref{fig:CBSSPOL_all_f} along
with the current exclusion regions. As we can see, for the wide
region of the ALP mass, $m_a = 10 \mathrm{\ GeV} - 5000$ GeV, our
CLIC bounds are much stronger than the bounds for the ALP production
in the LBL scattering at the LHC. They are also stronger than all
other exclusion regions for $m_a > 80$ GeV, except for a very small
area in between $m_a = 600$ GeV and $m_a = 900$ GeV, see
Fig.~\ref{fig:CBSSPOL_all_f}. By comparing our results on the
polarized LBL scattering with the unpolarized case, we can conclude
that the third energy stage of the CLIC with the polarized electron
beams have the greater physical potential to search for heavy ALPs,
especially in the ALP mass region 1000 GeV -- 2000 GeV.




\end{document}